\documentclass[sigconf, 9pt]{acmart}

\renewcommand\footnotetextcopyrightpermission[1]{}

\usepackage{fancyhdr}
\usepackage{braket}
\usepackage{xcolor}
\usepackage{color,soul}
\usepackage{caption}
\usepackage{subcaption}
\usepackage{graphicx}

\AtBeginDocument{%
  }

\setcopyright{acmcopyright}
\copyrightyear{2022}
\acmYear{2022}
\acmDOI{XXXXXXX.XXXXXXX}

\acmConference[ICCAD '22]{International Conference on Computer-Aided Design 2022}{Oct 29 -- Nov 03,
  2022}{San~Diego, CA}
\acmPrice{15.00}
\acmISBN{978-1-4503-XXXX-X/18/06}

\begin{document}
\fancyhead{}

\title{Quantum Machine Learning for Material Synthesis and Hardware Security (Invited Paper)}


\author{Collin Beaudoin}
\authornote{Both authors contributed equally to this research.}
\affiliation{%
  \institution{The Pennsylvania State University}
  \city{University Park}
  \state{PA}
  \country{USA}
}
\email{cpb5867@psu.edu}

\author{Satwik Kundu}
\authornotemark[1]
\affiliation{%
  \institution{The Pennsylvania State University}
  \city{University Park}
  \state{PA}
  \country{USA}
}
\email{satwik@psu.edu}

\author{Rasit Onur Topaloglu}
\affiliation{%
 \institution{IBM Corporation}
 \city{Hopewell Junction}
 \state{NY}
 \country{USA}
}
\email{rasit@us.ibm.com}

\author{Swaroop Ghosh}
\affiliation{%
  \institution{The Pennsylvania State University}
  \city{University Park}
  \state{PA}
  \country{USA}
}
\email{szg212@psu.edu}

\renewcommand{\shortauthors}{Beaudoin et al.}

\begin{abstract}
  Using quantum computing, this paper addresses two  scientifically-pressing and day to day-relevant problems,  
  %that affect day-to-day life of billions of people 
  namely, chemical retrosynthesis which is an important step in drug/material discovery and security of semiconductor supply chain. 
  %move to intro--Chemical retrosynthesis is the process of decomposing an initial target molecule into molecules that could create the target. Retrosynthesis requires billions of dollars and years of work, making it the perfect candidate for machine learning (ML). Unfortunately, ML models require millions of parameters to generate a potential solution, making large-scale retrosynthesis using classical machines an improbable task. The size requirements of the models lend themselves to quantum due to it's ability to represent an exponentially growing amount of information with a linear growth in qubit count. 
  We show that Quantum Long Short-Term Memory (QLSTM) is a viable tool for retrosynthesis. %We compare our results to classical LSTM to build a direct comparison to state-of-the-art classical approaches. 
  %We implement four different retrosynthesis methods to build a comprehensive understanding of current QLSTM abilities on NISQ era machines. 
  We achieve 65\% training accuracy with QLSTM whereas classical LSTM can achieve 100\%. However, in testing we achieve 80\% accuracy with the QLSTM while classical LSTM peaks at only 70\% accuracy!
%Mopve to intro- Integrated Circuits (ICs) suffer from a larger variety of threats such as manufacturing defects, counterfeiting, reverse engineering etc. Hardware Trojans (HTs), tamper the circuitry posing a threat to ICs trustworthiness since it could severely disrupt the system's functionality/security. With the growing interest in Quantum Machine Learning (QML) and its possible future applications, 
We also demonstrate an application of Quantum Neural Network (QNN) in the hardware security domain, specifically in Hardware Trojan (HT) detection using a set of power and area Trojan features. 
%We used a dimensionality reduction algorithm to reduce the feature count and trained a simple 2-qubit QNN model using the reduced feature-set. 
The QNN model achieves detection accuracy as high as 97.27\%.
\end{abstract}




\maketitle

\section{Introduction}

\begin{figure}[b]
    \vspace{-2mm}
  \centering
  \includegraphics[width=0.48 \textwidth]{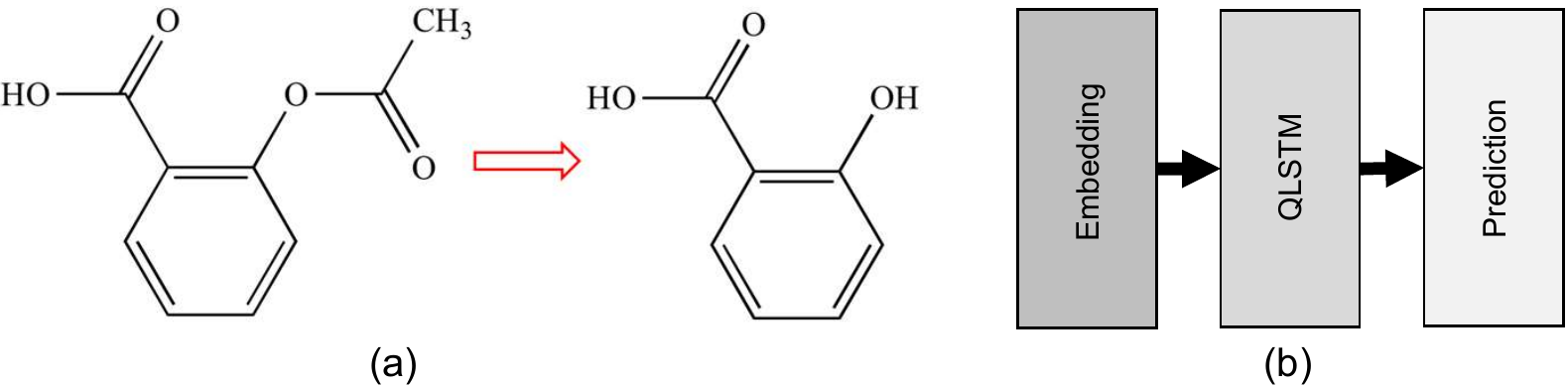}
  \vspace{-4mm}
  \caption{(a) Retrosynthesis example. Starting with a final molecule, the goal is the identify its starting molecule; (b) chemical retrosynthesis architecture used for training; embedding step turns information into the proper dimension for the QLSTM; the QLSTM learns and processes the data; the prediction step performs a softmax to convert the dimensional data to a singular value.} \label{fig:retro-intro}
  \vspace{-4mm}
\end{figure}

\textbf{Problem 1, chemical retrosynthesis:} Chemical retrosynthesis attempts to provide reactants that can be combined, using a chemical reaction, to synthesize a desired molecule. This process defines fields such as agriculture, medical treatment, material discovery, and countless others. Fig. ~\ref{fig:retro-intro}a exemplifies the retrosynthesis process, where the chemical on the left can be formed by the chemical on the right in combination of a chemical reaction. Performing retrosynthesis in the lab using trial-and-error takes years, and possibly cost billions of dollars, to resolve just for a single chemical. This leads to an immense amount of interest in machine learning (ML)-based solutions. Previous work have been able to generate promising results, but suffer from limitations. For example, expert defined rules for retrosynthesis \cite{socorro2005robia} relies on human's incomplete knowledge of retrosynthesis and doesn't scale well as more rules are being defined. To overcome limitations of domain knowledge, models have been created that do not require prior knowledge \cite{liu2017retrosynthetic, wang2021retroprime}. These solutions 
%are problematic for other reasons, they 
ignore the certainty of domain knowledge, require excessive training time,
%, while still having issues with producing chemically possible solutions. Worse yet are
and still poses scalability issues, making it hard to solve retrosynthesis of large molecules \cite{dai2019retrosynthesis}. Another common issue is a dependence on a predefined library of solutions rather than coming up with unique chemical results \cite{gomez2016design}. The efforts to resolve these issues run into the difficulty of finding chemically viable solutions, long training times, etc. \cite{gomez2018automatic,rappoport2014complex}. Chemical retrosynthesis could benefit from more capability than what modern machines offer, prompting us to search for solutions in new hardware domains.

The promise of exponential growth in computational space has led to the idea of Quantum Neural Networks (QNN) \cite{kak1995quantum} and more recently the Quantum Long Short-Term
Memory (QLSTM) \cite{chen2022quantum}. Unfortunately, Quantum Machine Learning (QML) efforts have fallen short of their desired exponential gain in speed \cite{ciliberto2018quantum}. However, they still offer the ability to represent an exponentially growing amount of information with only a linear growth in hardware size. %Focusing on the potential for the exponential ability of representation we aim at a QLSTM implementation. 

We evaluate the performance of QLSTM (a quantum-classical hybrid approach) and compare it to the performance of LSTM in its ability to make retrosynthetic predictions using the USPTO-50k dataset \cite{lowe2012extraction}. 
%In this paper we propose the novel use of QLSTM for chemical retrosynthesis. 
We also introduce two unique approaches to simplify the retrosynthesis process by identifying a specific substring within the reactants that are used to produce the given reaction. 

% \begin{figure}[b]
%   \centering
%   \includegraphics[width=0.5 \textwidth]{figs/arrows16.pdf}
%   \caption{Retrosynthesis example. Starting with a final molecule, the goal is the identify its starting molecule.} \label{fig:retro-ex}
%   \vspace{-4mm}
% \end{figure}

% \begin{figure}[!htb]
%     \vspace{-2mm}
%      \centering
%      \begin{subfigure}{0.46\textwidth}
%       \centering
%       \includegraphics[width=0.5 \textwidth]{figs/arrows16.pdf}
%       \caption{} \label{fig:retro-ex}
%       \vspace{1mm}
%      \end{subfigure}
%      \vspace{1mm}
%      \begin{subfigure}{0.46\textwidth}
%         \centering
%         \includegraphics[width=0.3\textwidth]{figs/retro-structure.pdf}
%         \caption{}   \label{fig:retro-arch}
%         \vspace{-2mm}
%      \end{subfigure}
%      \vspace{-2mm}
%         \caption{(a) Retrosynthesis example. Starting with a final molecule, the goal is the identify its starting molecule. 
%         (b) Chemical retrosynthesis architecture used for training; embedding turns information into the proper dimension for the QLSTM; the QLSTM learns and processes the data; the prediction performs a softmax to convert the dimensional data to a singular value.
%         }
%         \label{fig:retro-intro}
%     \vspace{-2mm}
% \end{figure}

\textbf{Problem 2, security of semiconductor supply chain:} In recent years, the hardware supply chain has been flooded with low-quality counterfeit Integrated Circuits (IC). The ICs suffer from a variety of threats/vulnerabilities such as, manufacturing defects, malicious circuitry, reverse engineering, etc. Hardware Trojans (HTs) tamper the circuitry posing a threat to IC trustworthiness since it could severely disrupt system functionality/security. Prior work has exploited classical ML to automate and provide a more reliable solution to the HT detection problem. In \cite{hasegawa2017Trojan}, 51 Trojan features are proposed to describe Trojan nets from gate-level netlists, use a random forest classifier to extract the best 11 Trojan features, and train a classifier to perform the classification task. 
%Similarly, in \cite{han2019hardware}, Trojan circuit features are extracted from RTL source codes and used a gradient boosting algorithm based model as a classifier for Trojan detection. 
The work in \cite{yasaei2021gnn4tj} utilizes a graph data structure for hardware representation and generates Data Flow Graphs (DFG) from RTL codes. Then Graph Neural Network (GNN) is used to extract features from the DFG and detect the presence of HT. A possible application of QML in classifying PCB defects (which can severely hinder system performance if undetected) from images is proposed in \cite{kundu2022security}. However, detection of HT has not been addressed yet.
%\hl{RT: is this statement really correct, please review? It seems incorrect. Do you mean in quantum domain only?}\hlgray{ SK: Yes, meant to say in quantum domain}

 To solve the above challenge, we evaluate the performance of a QNN model in detecting HTs from a small number of features. We also compare the results of our QNN with a few traditional ML classifiers and neural networks. Specifically, we use a Trojan feature dataset consisting of 50 features (of area and power), reduce feature size to 2 features using a dimensionality reduction algorithm, T-distributed Stochastic Neighbor Embedding (t-SNE), and then train a 2-qubit QNN using those features to evaluate the performance of the quantum model.

The rest of the paper is structured as follows: we cover basics on quantum computing, QLSTM, QNN etc. in Section \ref{background}, discuss the methodology used for chemical synthesis and Trojan detection in Section \ref{methodology}, present the results of both problems in Section \ref{results}, and end with closing remarks in Section \ref{conclusion}.

\section{Background} \label{background}
\text
% \textcolor{red}{inital para shpuld be reworded or removed}
% Since the inception of quantum computers it has been a race to create usable quantum hardware, and now that we have successfully created Noisy Intermediate-Scale Quantum (NISQ) machines \cite{preskill2018quantum, arute2019quantum} we aim to explore potential applications of our current and near future hardware. One of these applications is the fast paced field of machine learning. More specifically, we are exploring the recent works of QLSTM \cite{chen2022quantum, di2022dawn} in hope of finding a new application of the tool. To help understand why we want to use a quantum machine learning, and the benefits of quantum we discuss two important pieces: qubits and quantum gates. Following this discussion we introduce previous works on material discovery, and address our reason for use of QLSTM.

\subsection{Material Discovery}

% \textcolor{red}{always use present tense like -- is or has been but not "was"}
Material discovery extensively employs USPTO-50K dataset \cite{lowe2012extraction} which consists of 40,000 training, 5,000 validation, and 5,000 testing SMILES formatted chemical examples. SMILES is originally created as a way to use characters to represent chemical chains \cite{weininger1988smiles}. The letters represent various elements within the chain where the first letter of an element can be uppercase, denoting that the element is non-aromatic, or lowercase, denoting that the element is aromatic. If an element requires a second letter it will be lowercase, regardless of the casing of the first letter. Numbers are used within the chain to represent the opening and closing of a ring. Finally, parenthesis are used to denote branches from a chain, whereas periods are used to denote the start of a new chemical. %\hl{RT: what about numbers, what about lowercase vs. uppercase?} \hlgray{CB: resolved}  

\vspace{-2mm}
\begin{equation} \label{eq:1}
{<}RX\_1{>} c 1 c c c ( C n 2 c c c 3 c c c c c 3 2 ) c c 1
\end{equation}

The input from USPTO-50K consists of two parts, the first part is the reaction type that causes the reaction whereas the second part of the string is the reaction. The reaction type consists of 10 different possible values, ranging from 1-10. The output consists of possible input reactants that can be used in combination with the reaction type to create the final reaction. Exemplifying the SMILES format in Eq. ~\ref{eq:1}, the initial six characters, (${<}RX\_1{>}$), represent the reaction type that causes the target molecule given certain reactant(s). Following the reaction type, we have the chemical $c 1 c c c ( C n 2 c c c 3 c c c c c 3 2 ) c c 1$, which breaks down into three unique pieces. $c 1 c c c$ makes the initial chain, while $( C n 2 c c c 3 c c c c c 3 2 )$ forms a separate chain, which is denoted by the parentheses. Finally, we end with a third smaller chain, $c c 1$. Next is the use of $C$ and $c$, in the uppercase we note there is only a single non-aromatic carbon used, while the rest of the carbon in the chain is aromatic. Finally, we consider the use of numbers. Within the separate chain marked by the parentheses, we note the smallest ring formed, $ 3 c c c c c 3$, this is the third ring in the set, which is why it is marked by two different 3s. Since the creation of the USPTO-50K, it has frequently been used as an experimental testing ground for chemical retrosynthesis \cite{shi2020graph, dai2019retrosynthesis, wang2021retroprime}. We note that due to the nature of this difficult problem and unlike familiar benchmarks in other domains, the accuracy of much of this work rarely reaches higher than 50\% while predicting the proper reactant for a given input. Although 50\% is typically associated with random guessing, in this domain the accuracy relies on the exact match of reactant(s) to the given reaction. Given each reaction can have one or two reactants, and the majority of these reactants are unique to their reaction, it is easy to see 50\% accuracy is far higher than random guess. %\hl{RT: Please explain why 50\% is not just random guess} \hlgray{CB: I added a description above, let me know if I need to adjust}. 
We summarize the results from previous work in Table ~\ref{tab:1}.

\begin{table}[t]
\vspace{-4mm}
  \caption{Small summary of SOTA chemical retrosynthesis results.}
  \vspace{-2mm}
  \label{tab:1}
  \begin{center}
        \begin{tabular}{|c|c|}
            \hline
            \textbf{Model Type} & \textbf{Resulting Top 1 Accuracy}\\
            \hline
            G2Gs \cite{shi2020graph} & 48.9\% \\
            \hline
            GLN \cite{dai2019retrosynthesis} & 52.5\% \\
            \hline
            RetroPrime \cite{wang2021retroprime} & 51.4\% \\
            \hline
            Augmented Transformer \cite{tetko2020state} & 53.5\% \\
            \hline
        \end{tabular}
  \end{center}
  \vspace{-4mm}
\end{table}

\subsection{\textbf{Qubits}}

The qubit is the basis for all quantum computing, similar to its classical counterpart, the bit. But, there is a significant advantage of the qubit. Unlike the classical bit a qubit stores a mix of two states together, which is called superposition. For a single qubit, the states $\ket{0} = \begin{bmatrix} 1 \\ 0 \end{bmatrix}$ and $\ket{1} = \begin{bmatrix} 0 \\ 1 \end{bmatrix}$ are called our basis states. It is from these basis states that almost all quantum computation stems from.

\subsection{\textbf{Quantum Gates}}

Quantum gates are operations that are performed on qubits, similar to classical gates. These quantum gates are used to change the state of the qubits on which the operation is being performed. They typically are represented in the form of unitary matrices which operate on some initial qubit state. The most common quantum gates are the Hadamard (H), Bit flip (X) and Rotation gate (RX, RY, RZ) which are all single qubit gates. While the Controlled Not (CNOT) is a two-qubit gate. These gates allow us to perform almost all of our basic encodings of data in the quantum state, allowing for meaningful computation of quantum information.

\subsection{\textbf{LSTM}}

LSTM is an adaptation of the original Recurrent Neural Network (RNN) structure which is designed to keep temporal storage of information. This allows the neural network to maintain previous states of information. However, there is no guarantee as to what information is held and for how long it will remain, causing saturation issues. To get around these issues the LSTM allows for the neural network to decide when to add/remove pieces of information, helping mitigate context saturation issues.

% \begin{figure}[t]
% \vspace{-2mm}
%   \centering
%   \includegraphics[width=0.3\textwidth]{figs/retro-structure.pdf}
%   \caption{Chemical Retrosynthesis architecture used for training; embedding turns information into the proper dimension for the QLSTM; the QLSTM learns and processes the data; the prediction performs a softmax to convert the dimensional data to a singular value }   \label{fig:retro-arch}
%   \vspace{-2mm}
% \end{figure}

\begin{figure}[t]
\vspace{-2mm}
  \centering
  \includegraphics[width=0.48\textwidth]{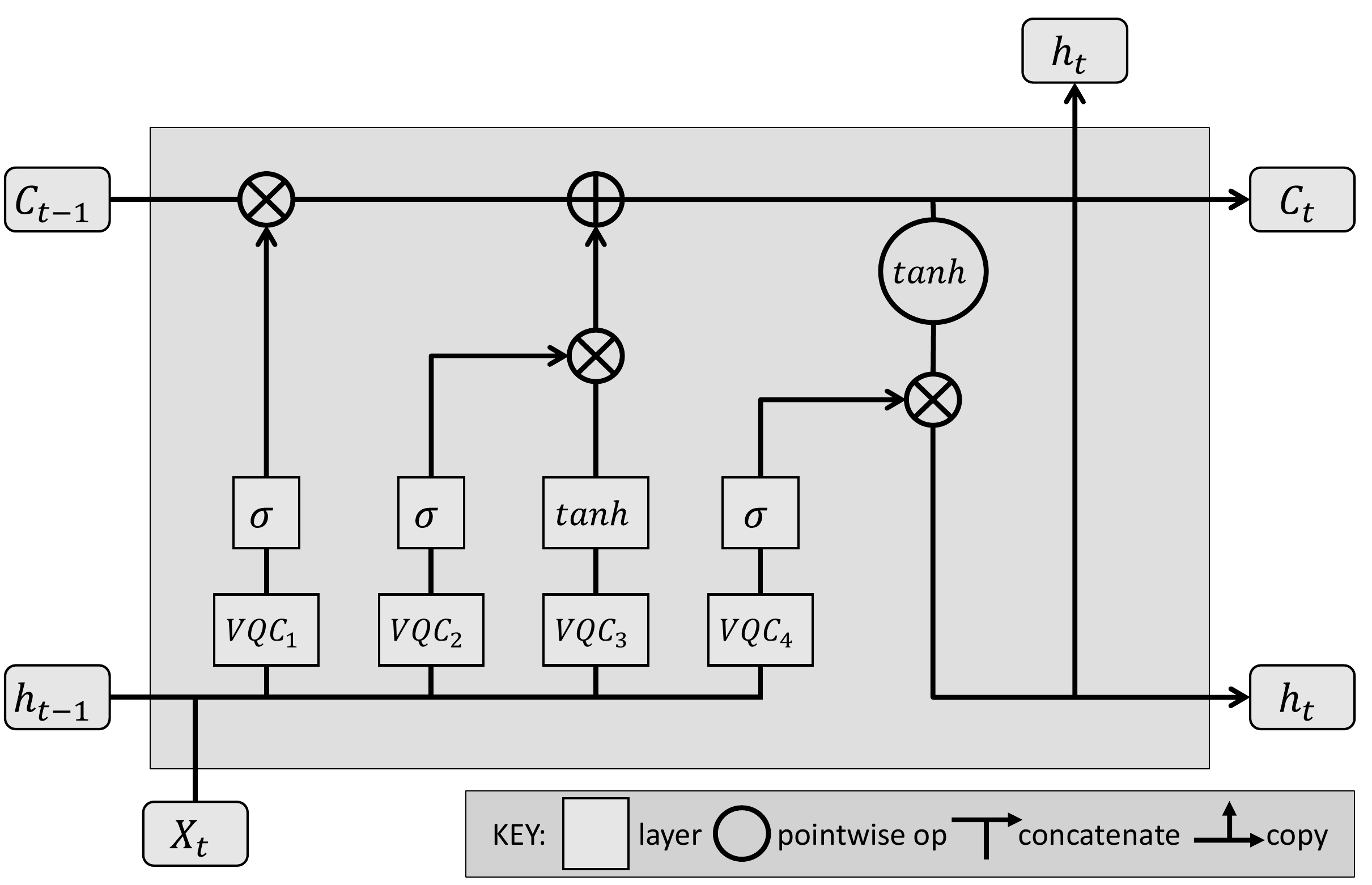}
  \vspace{-5mm}
  \caption{QLSTM architecture used for training; ($C_{t-1}, C_t$) represent the cell state, ($X_{t}$) represents the input, ($h_{t-1}, h_t$) represent the output state. The $VQC_1$ wire represents the forget gate, deciding if the input should be added to memory. The ($VQC_2, VQC_3$) wires represent the update gate, updating the cell memory if there is need. The $VQC_4$ wire represents the output gate, outputting the result of the QLSTM to the rest of the model. }   \label{fig:QLSTM}
  \vspace{-6mm}
\end{figure}

\subsection{QLSTM}

There have been many attempts in the quantum computing domain to create trainable networks \cite{cong2019quantum, kak1995quantum} to solve classification problems. However, selective memory has not been available.
QLSTM addresses this challenge and offers the same advantage as classical LSTMs, i.e., the ability to intentionally form a contextual understanding of previous input. This approach is near identical to the classical LSTM. The divergence of the two occurs when the network, instead of taking the information directly from the hidden layer and the input, takes the information and pass it to a Variational Quantum Circuit (VQC) where we can perform a data entanglement of the values. We then perform a measurement on the entangled information and proceed to process it in the same prediction structure as the classical LSTM. Fig. ~\ref{fig:retro-intro}b displays the basic overall architecture of the network: embedding, QLSTM and prediction. Embedding is preferred to a bag of words model as it reduces potentially large sparse vectors to smaller dense vectors that require less memory. Fig. ~\ref{fig:QLSTM} shows the structure of the QLSTM. 
%\hl{RT: Would like to see some improvements to the figure. Round shapes touching Ct-1 needs cleaned up, tanh should fit without touching border. Why is one tanh in circle and the other in rectangle? Can you label the arrow that goes up? So all lines represent 4 qubits? If VQC is shown in Figure 3, then where does classical features come from to Figure2? How do the cross and plus operators work exactly? What is sigma} \hlgray{CB: Updated image and description} 
Starting with our representation choices, we use the round edge boxes to represent the external values fed into the QLSTM, each varying in size. For $X_t$ the size is dependant on the embedding layer whereas for $h_t$ and $C_t$ the size is defined by the hidden dimension size. We use the sharp-cornered boxes to represent layers of a network, and circles to represent pointwise functions. For the wires, as displayed in the key, we use the wire merging to represent concatenation and wire splitting to represent a copy of the wire. We also use \(\sigma\) to represent the sigmoid activation function, defined by Eq.~\ref{eq:sigmoid-fxn}, and tanh is the arctan activation function,  defined by Eq.~\ref{eq:arctan-fxn}.

\begin{equation} \label{eq:sigmoid-fxn}
\sigma(x) = \frac{1}{1 + e^{-x}}
\end{equation}

\begin{equation} \label{eq:arctan-fxn}
tanh(x) = \frac{2}{1 + e^{-2x}} - 1
\end{equation}

\begin{figure}[t]
\vspace{-2mm}
  \centering
  \includegraphics[width=0.45\textwidth]{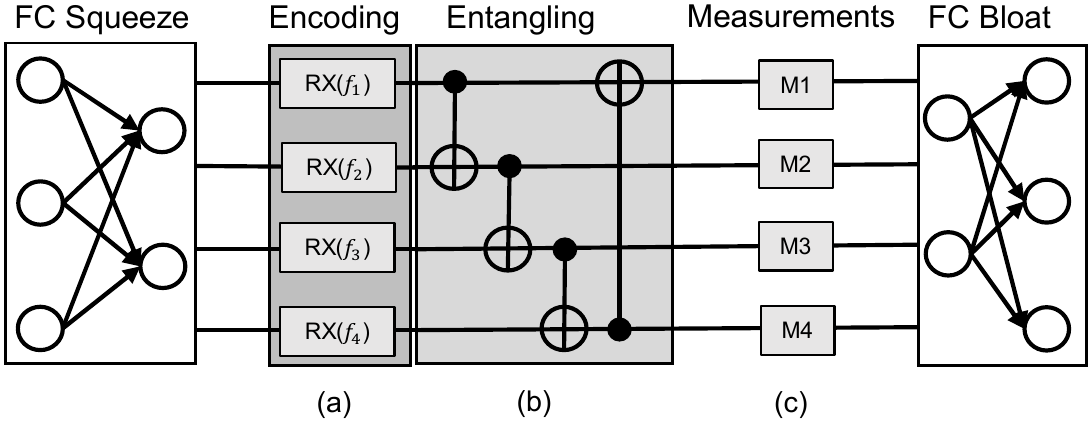}
  \vspace{-2mm}
  \caption{Modified basic entangler circuit; fully connected (FC) squeeze layer reduces the input size to be of the same qubit count. (a) Angle encoding converts classical features ($f_1, f_4, f_3, f_4$) to quantum states, (b) parametric quantum circuit entangles quantum states, (c) qubits measured, and bloated to original higher-dimension space. \cite{bergholm2018pennylane}.} \label{fig:basic_entangler}
  \vspace{-9mm}
\end{figure}

Working through the QLSTM starting with the bottom left, we have \(X_{t}\) which represents the input to the QLSTM structure. The input is concatenated with the previous hidden layer information, which is represented as \(h_{t-1}\). This combination is fed into four different VQCs; each of them are defined by a modified version of the basic entangler circuit from Fig. ~\ref{fig:basic_entangler}.
% \textcolor{red}{(reduce size of the fig and remove background color)}.

The modified basic entangler includes a trainable fully connected layer that squeezes the dimensional space of the information down to the circuit size of the VQC. After each VQC a quantum measurement of the expectation for each wire is fed to the trainable fully connected bloating layer. The bloating layer, increases the size from the quantum circuit back to the required dimensional space of the classical network. This is then processed using classical LSTM approaches. Hence, this is a quantum-classical hybrid approach. The first sigmoid activation is known as the forget gate which is used to decide whether to update the context \(C_{t}\) to include the new input. After the sigmoid, the result is multiplied onto \(C_{t - 1}\).  The second sigmoid, and the \(tanh\) activations are  known as the input gate which is used to write the new input into the context. The result of the sigmoid and tanh activations are multiplied to either be added to \(C_{t - 1}\), or to ensure the input is not added to \(C_{t - 1}\). The last sigmoid activation is known as the output gate where the actual prediction is performed. This output is also used to update the hidden layer \(h_{t}\).

\subsection{QNN}
QNN is a promising QML model that has received a lot of attention in recent years. A traditional QNN is made up of a data encoding circuit, a Parametric Quantum Circuit (PQC), and measurement operations. The data encoder transforms classical data into a quantum state. The PQC transforms the quantum state using a chosen ansatz.
%\hl{RT: transforms how exactly?} \hlgray{Explained in the next paragraph}. 
Measurements determine the output state. The PQC parameters are tuned during the training phase to produce the desired measurement results. We can train QNN models to perform traditional ML tasks such as classification, regression, distribution generation, etc. by selecting appropriate cost functions.

\begin{figure}[t]
\vspace{-2mm}
  \centering
  \includegraphics[width=0.45\textwidth]{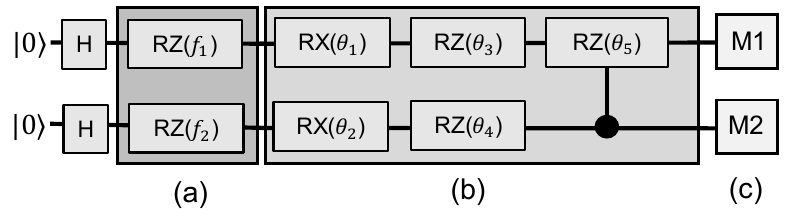}
  \vspace{-3mm}
  \caption{QNN architecture used for training; (a) angle encoding to convert classical feature ($f_1, f_2$) to their corresponding quantum state; (b) parametric quantum circuit used to perform desired transformations; and (c) measurement operation which collapses the qubit state to 0 or 1.}
  \label{fig:qnn}
  \vspace{-4mm}
\end{figure}

Fig. \ref{fig:qnn} shows the architecture of the 2-qubit QNN we used for training. It consists of the (a) Encoding, (b) PQC and (c) Measurement blocks. Several encoding techniques have been explored, e.g., amplitude encoding, basis encoding, NEQR \cite{zhang2013neqr}. We employ angle encoding where we pass classical features ($f_1, f_2$) as angles of quantum rotations gates ($RZ$) to transform them to quantum state. Similar to angle encoding, there are a number of PQC ansatz \cite{sim2019expressibility} to choose from but almost all of the PQCs consists of two main gate types: single qubit gates which are used to perform design space exploration, and two qubit gates which are used to entangle the qubits. The latter forms a correlation between the qubits based on the input feature values.

In the QNN, we use the 2-qubit Controlled-RZ ($CRZ$) gate to entangle the qubits and rotation gates along X and Z axes ($RX, RZ$) for transformation/exploration. The PQC/QNN is analogous to a classical neural network where we adjust the weights ($w_i$) to reduce the loss value while we adjust the tunable parameters ($\theta_i$) to generate the desired output in QNN. Finally, to measure the qubit state the most widely used measurement technique is Pauli's measurement along any of the X, Y, or Z axes. In our QNN model we used Pauli-Z measurement ($\sigma_z = \begin{bmatrix} 1 & 0 \\ 0 & -1 \end{bmatrix}$). A measurement in the Pauli Z basis means projecting the state onto one of the states $\ket{0}$ or $\ket{1}$ (the eigenstates of Pauli Z matrix).

\section{Proposed Methodology} \label{methodology}
\subsection{Chemical Retrosynthesis} 
Previous works on retrosynthesis have addressed the problem from many different ways such as, using graph, transformer or some other approach. Prior to transformers, LSTMs were the preferred approach to neural networks that required a memory \cite{vaswani2017attention}. However, LSTMs don't work as well as transformers. Therefore, we propose two unique approaches to simplify the problem of retrosynthesis namely, (a) 
%\hl{RT: single reaction and first reaction type, are these two different things?} \hlgray{CB: Typo, thanks for catching} 
we restrict the reactions by selecting just a single reaction type, \({<}RX\_1{>}\), in an attempt to simplify the retrosynthesis process. This subset is reduced from 12,000 to just 9 samples to reduce training time, and emulate the proof of concept proposed by Di Sipio \cite{di2022dawn} %\hl {RT: how many samples did it start from?} \hlgray{CB: added the original sample count for first reaction type}.
(b) We revert back to including all reaction types and change our output from a prediction of the reactants to a prediction of a chemical chain within the reactant. For this we select acetic acid and acetone as the common chemical chains and reduce the input reactions to only options that produce the selected chemical chains. This subset is reduced from 2,100 samples down to 200, which is then splitted  90:10 between training-validation set so there are 180 training samples and 20 validation samples. We then introduce these approaches to the QLSTM to show the potential of quantum computing in chemical retrosynthesis. In order to implement the encoding and the required layers for both LSTM and QLSTM, as well as the sigmoid activation and arctan activation function for the QLSTM, we use Pytorch  \cite{NEURIPS2019_9015}. The quantum circuits are trained using pennylane \cite{bergholm2018pennylane}.

\subsection{Hardware Trojan Detection} 
Here we consider a 50-feature dataset \cite{liakos2019machine}, which was originally created from Trojan free (TF) and Trojan infected (TI) circuits/ benchmarks available in Trust-Hub \cite{shakya2017benchmarking}, a public benchmark library. 
%\hl{RT: who certifies it?} \hlgray{SK: not really certified, sponsored by NSF, removed it, I think I wrote it in a flow.}. 
This original feature set containing area and power characteristics of the TI/TF circuits has been created using an industrial circuit design tool (DC compiler Synopsys). However, the feature set had a total of $\sim$900 samples among which very few samples of TF circuits were present compared to TI ones with a TF:TI ratio of 1:40. Thus, a reproduction technique (e.g., by repeating the TF circuit features to match the number of TI ones for each circuit/benchmark category) %\hl{RT: which reproduction technique exactly?} 
has been used to balance out the ratio between TF and TI samples. %\hlgray{SK: The imbalance is caused by the fact that each trojan free circuit has several different trojan infected versions; thus, in order to balance out the dataset, trojan free circuit features are repeated to match the number of trojan infected ones (for each circuit/benchmark category).} 
The resultant feature set that we use for our evaluation purposes contains 3026 samples and 50 features. We tested our models on both the original and reproduced/augmented dataset. 

Since it is not ideal to directly train a QNN using a dataset containing such larger number of features, we compress the information down to a handful of meaningful features. 
%can make any ML algorithm computationally efficient. 
In the noisy quantum computing era, with access to a hardware with low qubits, it is critical to reduce dataset dimension to train QNNs efficiently. Although we can run quantum simulations in classical computers, they incur a very high computational cost. As a result, we use a non-linear dimensionality reduction technique, specifically T-distributed Stochastic Neighbor Embedding (t-SNE) \cite{van2008visualizing}, to reduce the feature size from 50 to 2 features for training our QNN. Although t-SNE is widely used as a visualization technique as it helps clearly visualize multiple class high dimensional data in 2D/3D space, it can also be used as a dimension reduction technique. This is true since it generates low number of high variance features which can help train networks/classifiers effectively. Lastly, we normalize the features %\hl{RT: can you formalize the normalization mathematically?} 
of this reduced dimension dataset before training our QNN (as shown in Fig. \ref{fig:qnn}). %\hlgray{SK: Explained here: }
More specifically, we use the "\textit{max}" normalization technique provided by the sklearn library \cite{scikit-learn}, which divides each feature value with the max feature value of that specific row ($x_{norm} = x/max(x)$). The need to normalize the features before training comes from the fact that, during the encoding step, as we are passing the features as rotation angle values of quantum gates, it is possible that feature values of different classes differ by a multiple of $2\pi$ and thus end up being treated as features of the same class by our QNN (as $RZ(f_1) = RZ(f_1 + 2\pi n)$).

\section{Results} \label{results}

\begin{figure*}[!htb]
    \vspace{-2mm}
     \centering
     \begin{subfigure}{0.4\textwidth}
         \centering
         \includegraphics[width=0.99\linewidth]{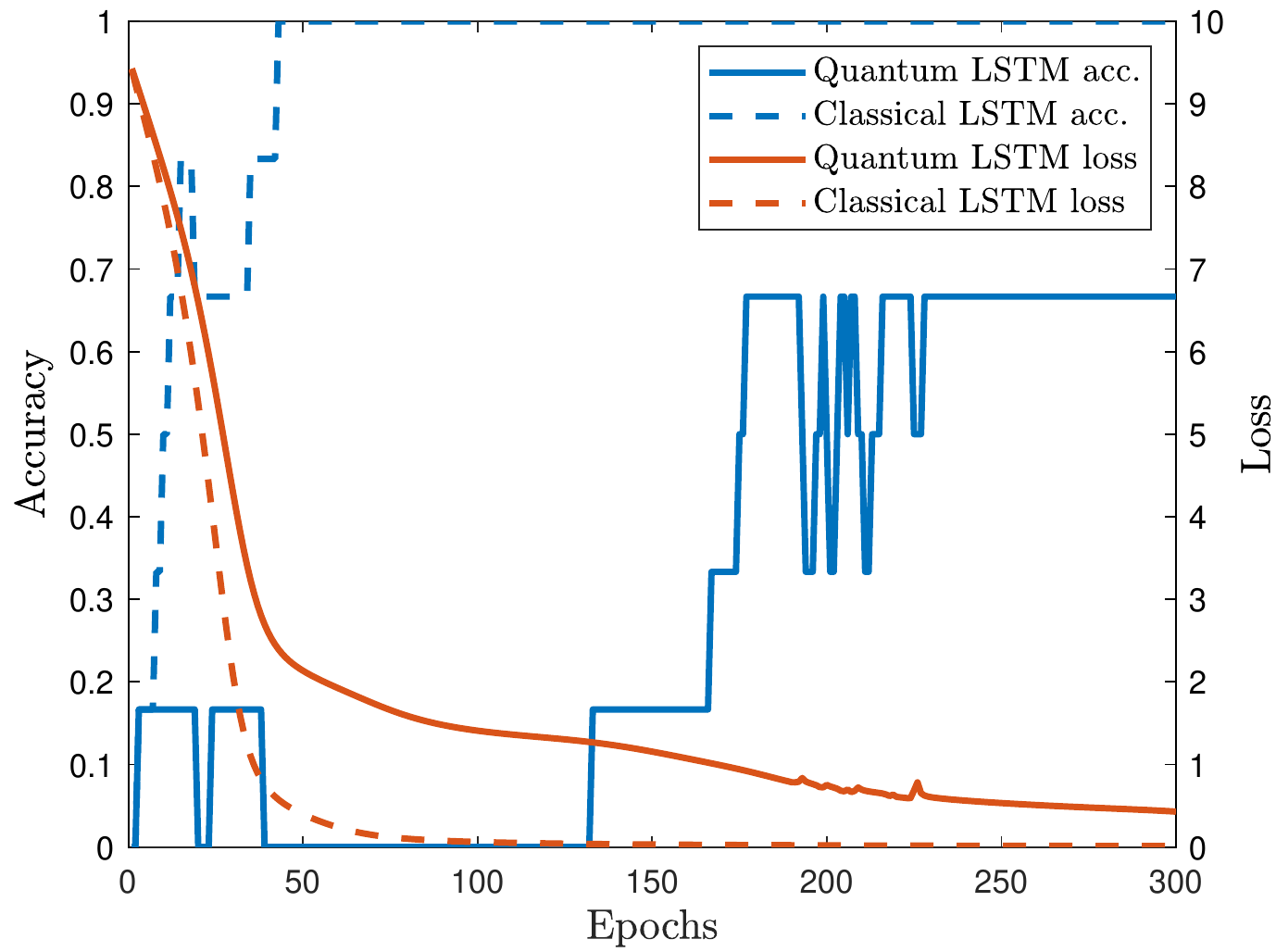}
         \vspace{-1mm}
         \caption{}
         \label{fig:sample-retro}
     \end{subfigure}
     \hspace{4mm}
     \begin{subfigure}{0.42\textwidth}
         \centering
         \includegraphics[width=0.99\linewidth]{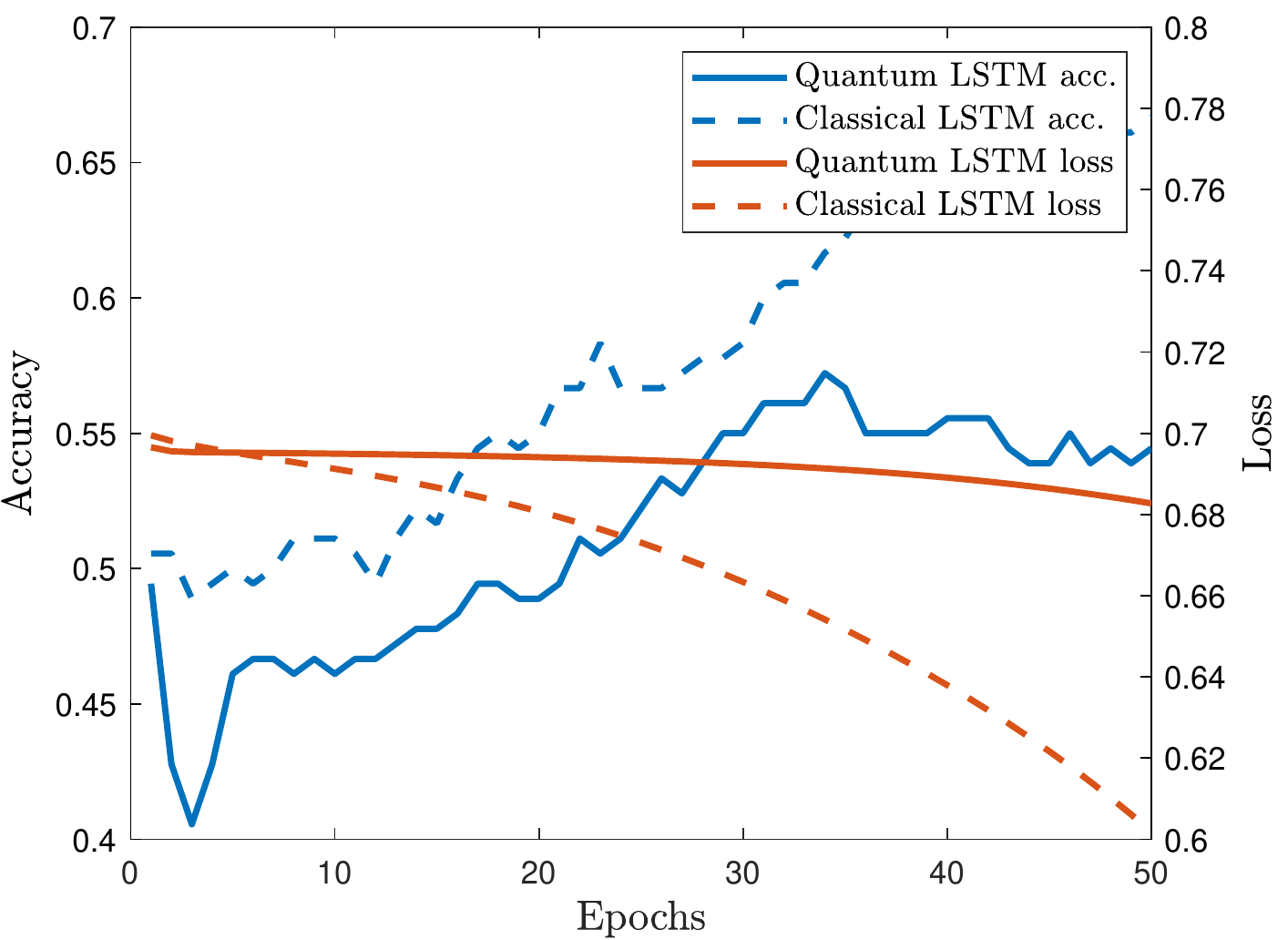}
         \vspace{-1mm}
         \caption{}
         \label{fig:chain-prediction}
     \end{subfigure}
     \vspace{-3mm}
        \caption{Results of chemical retrosynthesis using a quantum and classical LSTM model; (a) training of 9 chemical sample limited data set of a single reaction type (reaction type 1) where quantum is able to reach reasonable accuracy. 
        (b) training of 180 chemical sample limited data set of two common sub-string chemicals (acetic acid, acetone) where quantum nearly matches classical for the first 35 epochs.
        }
        \label{fig:smiles-results}
    \vspace{-4mm}
\end{figure*}

\subsection{Experimental Setups}

Since we adopt unique approaches to perform chemical retrosynthesis our results cannot be directly compared to other state-of-the-art work. For a fair comparison we create a classical LSTM in the same form as the QLSTM. The QLSTM depends on a 4 qubit VQC structure, while relatively small, the 4 qubit structure allows for a more manageable run time. Both the QLSTM and the LSTM use a small embedding dimension size of 8, and a small hidden dimension size of 6. The small embedding dimension is used for two reasons: first, it allows for enough memory for the second approach where we are predicting a reactant sub-string, and still uses a smaller vector than a bag of words would use for the proof of concept. The second reason is that when the concatenation of $X_t$ and $h_t$ occurs, it doesn't require a large fully connected layer squeeze/bloat to match the size of the VQC structure. The hidden dimension size is heuristically selected, using values less than the embedding dimension,  in expectation to keep the fully connected layer size requirement low. For the concatenation it is performed such that $X_t$ appends to $h_t$. Table~\ref{tab:parameters} contains a summation of parameters. All performance results are reported from execution on an Intel Xeon W-2125 CPU running at 4 GHz, with 16 GB of RAM. %\hl{RT: Did you utilize GPU acceleration?} \hlgray{CB: Not for training, so I removed}.

\begin{table}[h]
    \vspace{-2mm}
  \caption{Chemical retrosynthesis prediction defined parameters for both QLSTM and LSTM models}
  \label{tab:parameters}
  \vspace{-2mm}
  \begin{center}
        \begin{tabular}{|c|c|c|c|c|}
            \hline
            \textbf{Parameter} & $X_{t}$ & $h_{t}$ & $C_{t}$ & VQC size\\
            \hline
            \textbf{Value} & 8x1 dim & 6x1 dim & 6x1 dim & 4 qubits\\
            \hline
        \end{tabular}
  \end{center}
  \vspace{-2mm}
\end{table}

 For the second problem, the train-test split of 90:10 is used. We further used 10\% of the training data for validating the model %\hl{RT: Could you report exact numbers?} 
 Therefore, training, testing and validation employs 2452, 302 and 272 samples, respectively. We trained the QNN for 10 epochs with the following parameters; \textit{Loss function:} Sparse Categorical Cross Entropy \cite{tensorflow2015-whitepaper}, \textit{Optimizer:} Adagrad \cite{adagrad} %\hl{RT: need citations} \hlgray{SK: added necessary citations}
 , \textit{Learning rate:} 0.4, and \textit{Batch size:} 32. All simulations were performed using Pennylane's \cite{pennylane} \textit{default.qubit} device on a computer equipped with a 12th Gen Intel(R) Core(TM) i7-12700H and 16GB RAM. %\hl{RT: Did you utilize GPU acceleration?} \hlgray{SK: No, removed it}.

\subsection{Single Reaction Type Retrosynthesis}

Before beginning the single reaction type retrosynthesis some preprocessing of the original data is required. For clarity, an example string is provided for each step. To begin we take the initial input strings, as seen below. 
%\hl{RT: which file? Also need mention breaking up SMILES format in multiple lines does not mean anything special. } \hlgray{CB: Removed reference to file, and mentioned that the multiple lines don't mean anything}.

\vspace{-4mm}
\begin{multline*} 
{<}RX\_1{>} F \;c \; 1 \; c \; c \; 2 \; c \; ( \; N \; C \; 3 \; C \; C \; C \; C \; C \; C \; 3 \; ) \\ \; n \; c \; n \; c \; 2 \; c \; n \; 1
\end{multline*}

We note here that the use of multiple lines are meaningless, they are just inserted for readability purposes. For simplicity and legibility, we use superscript numbers to represent a repeating series of a character. This format will be followed for the continuation of the work:

\vspace{-2mm}
\begin{multline*}
{<}RX\_1{>} F \;c \; 1 \; c ^{2} \; 2 \; c \; ( \; N \; C \; 3 \; C \; ^{6} \; 3 \; ) \; n \; c \; n \; c \; 2 \; c \; n \; 1
\end{multline*}

As part of our method, we then compress by removing all of the individual spacing, as this space does not carry any special meaning in the context of SMILES format: %\hl{RT: however space around $RX_1$ is kept, please remove or mention why} \hlgray{CB: it was an issue with math representation in Latex....I found a workaround}

\begin{equation*} \label{eq:no_space_string}
{<}RX\_1{>} F c 1 c ^{2} 2 c ( N C 3 C ^{6} 3 ) n c n c 2 c n 1
\end{equation*}

We then ensure the reaction type is the first reaction type, matching the ${<}RX\_1{>}$. After this we strip off the reaction type as it is no longer helpful: 

\begin{equation*} \label{eq:no_rx_type_string}
F c 1 c ^{2} 2 c ( N C 3 C ^{6} 3 ) n c n c 2 c n 1
\end{equation*}

After finishing the input string, we take the output string for processing. Here we match the input string to the output string to find the corresponding output:
%\hl{RT: this index business is confusing; make a table and show some I/O pairs instead?. Also what does a dot mean? Can you remove the repetition between 4.1-4.5 and only talk about differences? Should 4.3 and 4.5 dropped because results are not good?} \hlgray{CB: Removed index, the dot description was added to the update of what numbers and casing meant, let me know if I should re-mention it here. 4.3 and 4.5 were dropped per email discussion}

\begin{equation*} \label{eq:result_string}
F \; c \; 1 \; c ^{2} \; 2 \; c \; ( \; C \; l \; ) \; n \; c \; n \; c \; 2 \; c \; n \; 1 \; . \; N \; C \; 1 \; C ^{6} \; 1
\end{equation*}

After we find the matching output string, we simply compress the string by removing the spaces:

\begin{equation*} \label{eq:no_space_result_string}
F c 1 c ^{2} 2 c ( C l ) n c n c 2 c n 1 . N C 1 C ^{6} 1
\end{equation*}

Once the trimming of the input and output is done, we perform a word encoding for both the input and the output to have a numerical representation of the SMILES strings for use in LSTM. The word encoding requires two unique lists, one for reactions and another for reactants. Each list consists of unique chemicals, where each chemical is assigned it's numerical value based on it's index within the list it belongs to. 
%\hl{RT: how exactly?} \hlgray{CB: Added description above} 
After completing the preprocessing we train the LSTM and QLSTM models. The promising results in Fig. ~\ref{fig:sample-retro} show that the quantum approach, while unable to match the results of classical approach, is able converge to an accuracy of 65\% and a loss of 0.1.

\begin{figure}[b]
\vspace{-4mm}
  \centering
  \includegraphics[width=0.4\textwidth]{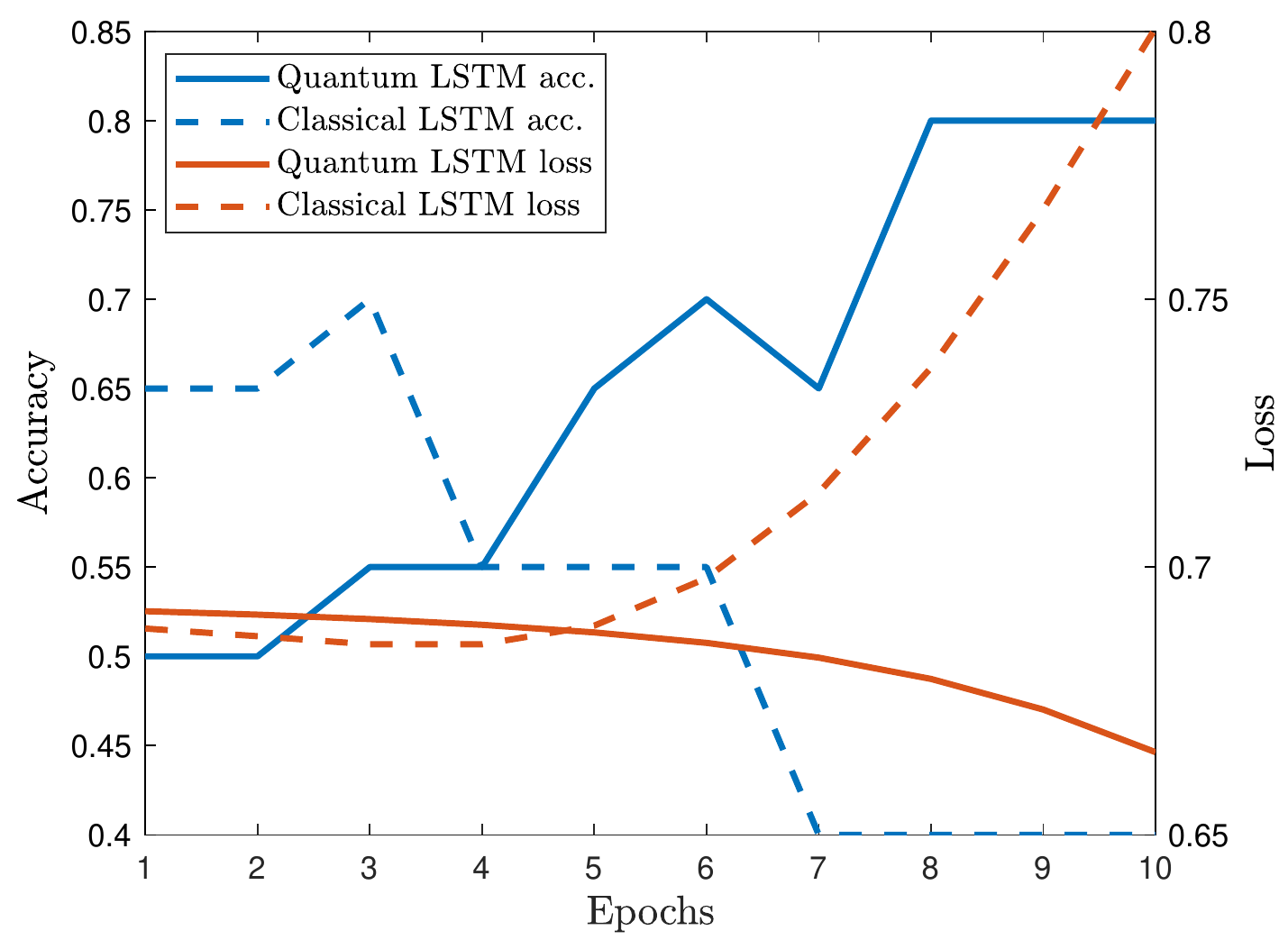}
  \vspace{-3mm}
  \caption{Results of 20 chemical sample limited data set testing of two common sub-string chemicals (acetic acid, acetone) where quantum outperforms classical.} %Validation is ran every 5 training epochs.} 
  \label{fig:chain-val}
  \vspace{-4mm}
\end{figure}

\subsection{Chemical Chain Prediction}

We perform similar preprocessing of the original data as explained with an example string below. We take the initial output strings from the input file:

\vspace{-2mm}
\begin{multline*}
C ^{2} \; ( \; C \; ) \; ( \; C \; ) \; O \; C \; ( \; = \; O \; ) \; N \; C ^{2} \; ( \; = \; O \; ) \; C ^{3} \; ( \; = \; O \; ) \\ \; O \; C ^{4} \; ( \; = \; O \; ) \; O \; C \; c \; 1 \; c ^{5} \; 1
\end{multline*}

We then remove all of the individual spacing:

\vspace{-2mm}
\begin{multline*} 
C ^{2} ( C ) ( C ) O C ( = O ) N C ^{2} ( = O ) C ^{3} ( = O ) O C ^{4} ( = O ) \\ O C c 1 c ^{5} 1
\end{multline*}

We then ensure the reaction contains the acetic acid chain, \(CC(=O)O\) or acetone \(CC(=O)C\). After this we dispose of the reactant and simply use the label of the chain the string contains, for example: \emph{acetic}.

%\begin{equation*} \label{eq:no_rx_type_string}
%acetic.
%\end{equation*}

After we finish the output string, we take the input string and match it to the output string to find the corresponding input:

\vspace{-2mm}
\begin{multline*} \label{eq:result_string}
{<}RX\_6{>} C ^{2} \; ( \; C \; ) \; ( \; C \; ) \; O \; C \; ( \; = \; O \; ) \; N \; C ^{2} \; ( \; = \; O \; ) \\ \; C ^{3} \; ( \; = \; O \; )  \; O \; C ^{4} \; ( \; = \; O \; ) \; O
\end{multline*}

After we find the matching input string, we remove the  spaces, and the reaction type:

\vspace{-2mm}
\begin{equation*} \label{eq:no_space_result_string}
C ^{2} ( C ) ( C ) O C ( = O ) N C ^{2} ( = O ) C ^{3} ( = O ) O C ^{4} ( = O ) O
\end{equation*}

Once the trimming of the input and output is done, we perform our encoding and we train the LSTM and QLSTM models. The results in Fig. ~\ref{fig:chain-prediction} show that the classical loss never reaches a point of convergence, where the quantum loss also doesn't reach convergence nor does it reach the same level as the classical. These results hold true for accuracy, where the classical domain reaches 65\% and the quantum domain reaches 55\%. While there is a small gap in performance, we see that given the task of identifying a common substring within the predicted reactants, quantum is able to nearly match classical performance during training. Validation is ran once every 5 epochs during training and here, there is a flip of performance. The results in Fig. ~\ref{fig:chain-val} show that the classical loss starts to increase after just 25 training epochs, whereas the quantum loss is steadily decreasing for the entirety of the training. As for the accuracy, the classical accuracy reaches a high of 70\% and steadily decreases to 40\%. For the quantum domain the accuracy starts at 50\%, while steadily increasing all the way to 80\%, outperforming the classical model by 40\% at the end of the model training.

\subsection{Trojan Detection} \label{Trojan:results:section}

\begin{figure}[t]
\vspace{-2mm}
  \centering
  \includegraphics[width=0.42\textwidth]{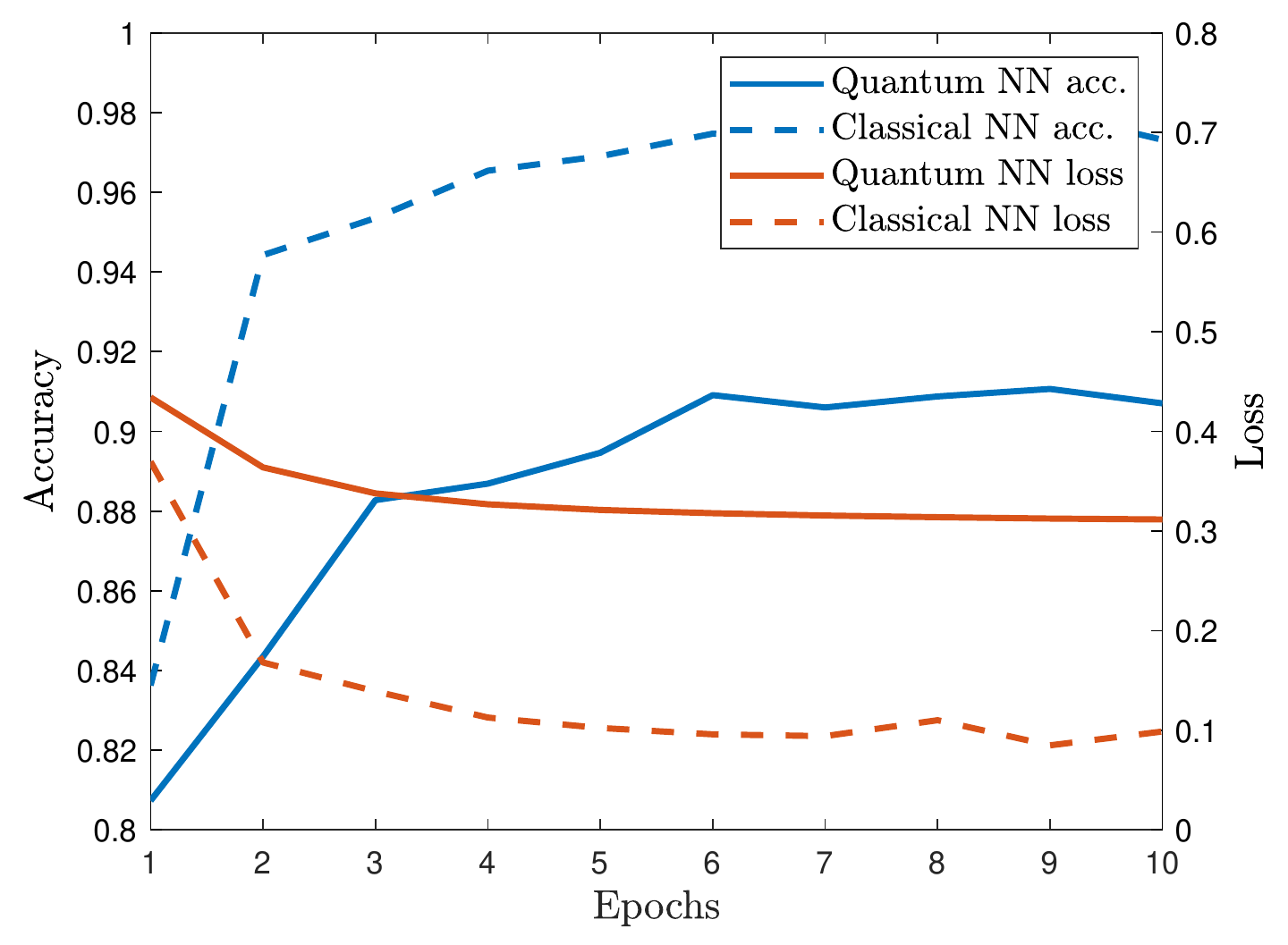}
  \vspace{-4mm}
  \caption{Results of Trojan detection using a quantum and classical neural network trained on augmented dataset of features $f = 2$.} 
  \label{fig:Trojan-plot}
  \vspace{-6mm}
\end{figure}

Coming to the second problem we study, Fig. \ref{fig:Trojan-plot} shows the accuracy and loss comparison of our QNN with a simple classical neural network on the augmented dataset with the following neuron configuration; 2-64-256-64-2 per the five layers from input to output. We trained our classical NN for 10 epochs [\textit{Optimizer:} Adam \cite{adam}, %\hl{RT: need citations} \hlgray{SK: added necessary citations} 
\textit{Loss\_Fn:} Sparse Categorical Cross Entropy \cite{tensorflow2015-whitepaper} and \textit{Learning rate:} 0.01]. The maximum training accuracy achieved by our QNN and classical NN was 91.06\% and 98.03\% respectively.

\begin{table}[b]
    \vspace{-4mm}
  \caption{Trojan detection accuracy of different models/classifiers on the augmented dataset of features $f = 2$.}
  \label{tab:accuracies}
  \vspace{-2mm}
  \begin{center}
        \begin{tabular}{|l|c|c|}
            \hline
            \textbf{Model} & \textbf{Training Acc.} & \textbf{Testing Acc.}\\
            \hline
            Perceptron & 73.72\% & 72.61\% \\
            \hline
            GaussianNB & 74.92\% & 76.73\%\\
            \hline
            LogisticRegression & 75.04\% & 73.76\% \\
            \hline
            \textit{QuantumNN} & \textit{91.06\%} & \textit{90.04\%} \\
            \hline
            SVM & 96.49\% & 96.04\% \\
            \hline
            ClassicalNN & 98.03\% & 98.35\% \\
            \hline
        \end{tabular}
  \end{center}
  \vspace{-4mm}
\end{table}

We also trained some of the linear/non-linear ML classifiers with the augmented dataset of features $f = 2$ and compared the results of the same with our QNN model. From Table \ref{tab:accuracies} we can see that QNN performs better than few of the linear/non-linear models (Perceptron/GaussianNB) but falls behind SVM and classical neural network. The results clearly show that the classes are not linearly separable because linear classifiers like Perceptron and Logistic Regression perform poorly, as shown in Table \ref{tab:accuracies}.

Without employing the reproducing technique, we also trained our QNN and traditional NN model with the original feature set ($\approx900$ samples). We only modified one parameter before training the models; the learning rate, which we lowered to 0.01 and 0.001 for QNN and classicalNN, respectively. In this case, the QNN model is found to be more effective at detecting HTs, with a classification accuracy of up to 97.27\%. The QNN model performed identical to classicalNN, which produced an accuracy of 97.09\%. As a result, we can conclude that QNN models can potentially perform similar to classical neural networks in some cases.

It should be noted that the goal of this work is not to demonstrate superior classification accuracy over classical counterparts, but rather to show a proof-of-concept application of QML in hardware security domain. We posit that further optimization of the feature count, qubits, layers, epochs and/or lower the learning rate could achieve higher detection accuracy.

\section{Conclusion and Future Work} \label{conclusion}

We have shown that QLSTM is a viable solution to solve chemical retrosynthesis problem, even with just 4 qubits. While QLSTM didn't train as well as its classical counterpart, it is able to reach a reasonable accuracy and loss metrics for the proof of concept. For example, quantum achieves 65\% accuracy and classical achieves 100\%. It again is able to reach a reasonable accuracy e.g., 55\% for quantum and 65\% for classical while attempting to predict substrings. 
%, where quantum reaches 60\% accuracy and classical reaches 65\% accuracy. 
However these gaps are misleading since quantum is able to reach an accuracy or 80\% whereas classical peaks at an accuracy of 70\% during testing of the substring prediction! 
%\hl{RT: report some numbers} \hlgray{CB: I added some numbers to show how quantum performs in comparison to classical} %However, by loosening our requirements from searching for a complete retrosynthetic solution to simply searching for a common chemical chain we note that QLSTM is able to perform nearly identical to classical. 
We also demonstrated a QNN application in hardware security domain, specifically Trojan detection from a set of area and power features. 
%generated using Synopsys tool. 
A very simple 2-qubit QNN with demonstrated ($\approx91\%$) accuracy is able to outperform some linear/non-linear classifiers which show $\approx75\%$ in terms of detection accuracy. In the future, the performance of the model can be improved by using a Quantum RAM (QRAM) to load the data and/or using a Quantum Graph Neural Network (QGNN) instead of QNN. 
%\hl{RT: report some numbers}
%on a graph depicting the hardware representation of Trojan infected circuits for Trojan feature extraction and detection, which could potentially improve detection accuracy even more.

\begin{acks}
The work is supported in parts by NSF (CNS-1722557, CNS-2129675, CCF-2210963, CCF-1718474, OIA-2040667, DGE-1723687, DGE-1821766 and DGE-2113839) and seed grants from Penn State ICDS and Huck Institute of the Life Sciences.
\end{acks}

\bibliographystyle{ACM-Reference-Format}
\bibliography{refs}

\end{document}